\documentclass[12pt]{article}

\usepackage{amssymb}
\usepackage{amsmath}

\begin{document} %

\title{\bf Physical models within the framework of the Randall-Sundrum scenario}

\author{A.V. Kisselev\thanks{
Electronic address: alexandre.kisselev@ihep.ru} \\
{\small Institute for High Energy Physics, 142281 Protvino, Russia}
\\
{\small and}
\\
{\small Theory Division, Physics Department, CERN,} \\
{\small 1211 Geneva, Switzerland} }

\date{}

\maketitle

\thispagestyle{empty}


\begin{abstract}
The Randall-Sundrum scenario with non-factorizable geometry and
fifth dimension $y$ being an orbifold, is studied. It has two branes
located at fixed points of the orbifold. The four-dimensional metric
is multiplied by a warp factor $\exp[\sigma(y)]$. Recently, a new
general expression for $\sigma(y)$ was derived which has the
orbifold symmetry $y \rightarrow - y$ and explicitly reproduces
jumps of its derivative on both branes. It is symmetric with respect
to the branes. The function $\sigma(y)$ is determined by the
Einstein-Hilbert's equations up to a constant $C$. In the present
paper we demonstrate that different values of $C$ result in quite
different physical schemes. Three schemes are considered, among
which are: (i) the RS1 model; (ii) the RSSC model with a small
curvature of the five-dimensional space-time; (iii) the
``symmetric'' scheme with $C=0$. The latter scenario is studied in
detail.
\end{abstract}




\section{Introduction}

The Randall-Sundrum (RS) scenario \cite{Randall:99} is a framework
with one extra dimension in a slice of the AdS$_5$ space-time
restricted by two 3-branes. Contrary to the ADD model
\cite{Arkani-Hamed:98}-\cite{Arkani-Hamed:99}, it solves the
hierarchy problem due to the higher-dimensional curvature rather
than the volume of the extra dimension.

The RS1 model \cite{Randall:99} predicts an existence of heavy
Kaluza-Klein (KK) excitations (massive graviton resonances) with the
lightest mass around few TeV. The phenomenological implications of
the RS1 model were explored in ref.~\cite{Davoudiasl:00} and next
publications. For the time being, graviton resonances are
intensively searched for at the LHC (see, for instance,
\cite{ATLAS:gravitons}, \cite{CMS:gravitons}).

Our main goal is to show that the RS scenario admits a variety of
models with different spectra of the KK gravitons, and,
consequently, diverse collider phenomenology. In fact, the RS1 model
is a particular physical scheme within the general framework of the
RS scenario. The interesting scheme (RSSC model) is realized when a
curvature of the five-dimensional space-time is much smaller than
the 5-dimensional Planck scale \cite{Giudice:05}-\cite{Kisselev:06}.
It predicts an alomost continuous spectra of the KK gravitons. In
the present Letter we study in detail the other scheme which was not
yet considered by other authors. All the schemes has quite different
experimental signature in searching for warped extra dimension.


\section{General solution for the warp factor}
\label{sec:2}

The RS scenario is described by the AdS$_5$ background warped metric
of the form
\begin{equation}\label{RS_background_metric}
\quad ds^2 = e^{-2 \sigma (y)} \, \eta_{\mu \nu} \, dx^{\mu} \,
dx^{\nu} - dy^2 \;,
\end{equation}
where $\eta_{\mu\nu}$ is the Minkowski tensor with the signature
$(+,-,-,-)$, and $y$ is the 5-th coordinate. It is assumed that the
periodicity condition $y = y \pm 2\pi r_c$ is imposed and the points
$(x_{\mu}, y)$ and $(x_{\mu}, -y)$ are identified. As a result, we
get the orbifold $S^1/Z_2$. After orbifolding, the coordinate $y$
varies within the limits $0 \leqslant  y \leqslant \pi r_c$. We
consider the scenario with two 3-branes located at the fixed points
$y = 0$ and $y = \pi r_c$. The SM fields are constrained to the
second of these branes (TeV brane).

The classical action of the Randall-Sundrum scenario
\cite{Randall:99} is given by
\begin{align}\label{action}
S &= \int \!\! d^4x \!\! \int \!\! dy \, \sqrt{G} \, (2 \bar{M}_5^3
\mathcal{R}
- \Lambda) \nonumber \\
&+ \int \!\! d^4x \sqrt{|g^{(1)}|} \, (\mathcal{L}_1 - \Lambda_1) +
\int \!\! d^4x \sqrt{|g^{(2)}|} \, (\mathcal{L}_2 - \Lambda_2) \;,
\end{align}
where $G_{MN}(x,y)$ is the 5-dimensional metric, with $M,N =
0,1,2,3,4$, $\mu = 0,1,2,3$. The quantities
\begin{equation}
g^{(1)}_{\mu\nu}(x) = G_{\mu\nu}(x, y=0) \;, \quad
g^{(2)}_{\mu\nu}(x) = G_{\mu\nu}(x, y=\pi r_c)
\end{equation}
are induced metrics on the branes, $\mathcal{L}_1$ and
$\mathcal{L}_2$ are brane Lagrangians, $G = \det(G_{MN})$, $g^{(i)}
= \det(g^{(i)}_{\mu\nu})$. Here and below the reduced mass scales
are used: $\bar{M}_{\mathrm{Pl}} = M_{\mathrm{Pl}} /\sqrt{8\pi}
\simeq 2.4\cdot 10^{18} \ \mathrm{GeV}$, $\bar{M}_5 = M_5
/(2\pi)^{1/3} \simeq 0.54 \, \bar{M}_5$.

From action \eqref{action} 5-dimensional Einstein-Hilbert's
equations follow
\begin{align}\label{H-E_equation}
\sqrt{|G|} & \left( \mathcal{R}_{MN} - \frac{1}{2} \, G_{MN}
\mathcal{R} \right) = - \frac{1}{4 \bar{M}_5^3} \Big[ \sqrt{|G|} \,
G_{MN}  \Lambda \nonumber \\
&+  \sqrt{|g^{(1)}|} \, g^{(1)}_{\mu\nu} \, \delta_M^\mu \,
\delta_N^\nu \, \delta(y) \, \Lambda_1 +  \sqrt{|g^{(2)}|} \,
g^{(2)}_{\mu\nu} \, \delta_M^\mu \, \delta_N^\nu \, \delta(y - \pi
r_c) \, \Lambda_2 \Big] \;.
\end{align}
For the 5-dimensional background metric
\eqref{RS_background_metric}, the Einstein-Hilbert's equations are
reduced to the following set of equations%
\footnote{Here and in what follows, the \emph{prime} denotes the
derivative with respect to variable $y$.}
\begin{align}
6 \sigma'^2 (y) &= - \frac{\Lambda}{4 \bar{M}_5^3} \;,
\label{sigma_deriv_eq} \\
3\sigma''(y) &= \frac{1}{4 \bar{M}_5^3} \, [\Lambda_1 \, \delta(y) +
\Lambda_2 \, \delta(\pi r_c - y)] \;. \label{sigma_2nd_deriv_eq}
\end{align}
As usual, we ignore the backreaction of the brane terms on the
space-time geometry.

The first solution of these equations was presented by Randall and
Sundrum \cite{Randall:99},
\begin{equation}\label{sigma_RS1}
\sigma_0(y) = \kappa |y| \;,
\end{equation}
with the fine tuning relations
\begin{align}\label{RS1_Lambda_fine_tuning}
\Lambda &= -24 \bar{M}_5^3\kappa^2 \;, \\
\Lambda_1 &= - \Lambda_2 = 24 \bar{M}_5^3 \kappa \;.
\label{RS1_Lambdas_fine_tuning}
\end{align}
The parameter $\kappa$ defines the magnitude of the 5-dimensional
scalar curvature $R^{(5)}$ in the region $0 < y < \pi r_c$, where
$R^{(5)} = - 20 \kappa^2$. The solution \eqref{sigma_RS1} is
consistent with the orbifold symmetry $y \rightarrow - y$. It is
\emph{not symmetric} with respect to the branes. That is why, the RS
solution \eqref{sigma_RS1} does not reproduce the jump of $\sigma'
(y)$ on the brane number 2 \emph{explicitly}, but it does it only
after taking into account periodicity condition.

Later on, an explicit expression which makes the jumps of
$\sigma'(y)$ on both branes was given in \cite{Dominici:03},
\begin{equation}\label{sigma_two_jumps}
\sigma(y) = \kappa \{y [2\,\theta(y) - 1] - 2 (y - \pi
r_c)\,\theta(y - \pi r_c) \}  + \mathrm{constant} \;.
\end{equation}
However, this expression is neither symmetric in variable $y$ nor
invariant with respect to an interchange of the branes.

Let us stress that both branes (or fixed points) must be treated on
an \emph{equal} footing. It means that instead of solution
\eqref{sigma_RS1} one can use the equivalent solution
\begin{equation}\label{sigma_RS_pi}
\sigma_\pi(y) = -\kappa |y - \pi r_c| + \kappa \pi r_c \;.
\end{equation}
Expression \eqref{sigma_RS_pi} is consistent with the orbifold
symmetry. Indeed, the periodicity condition means that the points
$y$ and $y \pm 2\pi r_c$ are identified. Consequently,
$\sigma_\pi(-y) = \sigma_\pi(y)$ due to the sequence of the
following equations
\begin{equation}\label{ofbifold_symmetry}
|-y - \pi r_c| \equiv |y + \pi r_c| = |y - \pi r_c| \;.
\end{equation}

The solution of eqs.~\eqref{sigma_deriv_eq},
\eqref{sigma_2nd_deriv_eq}, which realize the symmetry with respect
to the brane, is equal to $[\sigma_0(y) + \sigma_\pi(y)]/2$. Note
that both \eqref{sigma_RS1} and \eqref{sigma_RS_pi} are defined up
to an \emph{arbitrary constatnt}. As a result, we come to the
general solution of the Einstein-Hilbert's equations
\cite{Kisselev:13}-\cite{Kisselev:14}:
\begin{equation}\label{sigma_solution}
\sigma (y) = \frac{\kappa}{2} ( |y| - |y - \pi r_c | ) + C \;,
\end{equation}
where the constant $C$ can depend on a modulus of $\kappa$. The
5-dimensional and brane cosmological constants look like
\begin{align}\label{Lambda_fine_tuning}
\Lambda &= -24 \bar{M}_5^3\kappa^2 \;, \\
\Lambda_1 &= - \Lambda_2 = 12 \bar{M}_5^3 \kappa \;.
\label{Lambdas_fine_tuning}
\end{align}

This warp function \eqref{sigma_solution} has the following
properties: (i) it obeys the orbifold symmetry $y \rightarrow - y$;
(ii) the jumps of $\sigma'(y)$ are explicitly reproduced on both
branes; (iii) it is symmetric with respect to the branes. The latter
property is evident, since under the replacement $y \rightarrow \pi
r_c - y$, the positions of the branes are interchanged (the fixed
point $y=0$ becomes the fixed point $y=\pi r_c$, and vice versa),
while under the replacement $\kappa \rightarrow - \kappa$ the
tensions of the branes \eqref{Lambdas_fine_tuning} are interchanged.

An additional freedom of the RS scenario (constant $C$ in eq.
\eqref{sigma_solution}) will be used in the next section for model
building.

\section{Physical models in the RS scenario}
\label{sec:3}

In this section we will demonstrate that not only the brane warp
factors, but hierarchy relations and graviton mass spectra depend
drastically on a particular value of the constant $C$ in
eq.~\eqref{sigma_solution}. Correspondingly, the parameters of the
model, $\bar{M}_5$ and $\kappa$, differ significantly for different
$C$.

Let us define
\begin{equation}
\sigma_1 = \sigma(0) \;, \quad \sigma_2 = \sigma(\pi r_c) \;.
\label{sigmas_1_2}
\end{equation}
It follows from \eqref{sigma_solution} that
\begin{equation}\label{sigma_difference}
\Delta \sigma \equiv \sigma_2 - \sigma_1 = \kappa \pi r_c \;.
\end{equation}
The r.h.s of the hierarchy relations,
\begin{equation}\label{hierarchy_relation}
\bar{M}_{\mathrm{Pl}}^2  = \frac{\bar{M}_5^3}{\kappa} e^{-2\sigma_1}
\left( 1 - e^{-2\Delta \sigma} \right) \Big|_{\pi\!\kappa r_c \gg 1}
\simeq \frac{\bar{M}_5^3}{\kappa} e^{-2\sigma_1} \;,
\end{equation}
depends on the size of the extra dimension except for the case $C =
\kappa \pi r_c/2$.

There exist relations between physical parameters which look the
same for \emph{any} $C$. The masses of the KK graviton excitations
$h_{\mu\nu}^{(n)}$ on the TeV brane are defined from the equation
\begin{equation}\label{masses_eq}
J_1 (a_{1n}) Y_1(a_{2n}) - Y_1 (a_{1n}) J_1(a_{2n}) = 0 \;, \quad n
= 1, 2, \ldots \;,
\end{equation}
where $J_\nu(x)$, $Y_\nu(x)$ are Bessel functions, and $a_{in} =
(m_n/\kappa) \exp(\sigma_i)$. As a result, we get
\begin{equation}\label{m_n}
m_n = x_n \! \left( \frac{\kappa}{\bar{M}_5} \right)^{3/2} \!\!
\frac{\bar{M}_{\mathrm{Pl}}}{\sqrt{e^{2\Delta \sigma} - 1}}
\Big|_{\pi\!\kappa r_c \gg 1} \simeq x_n \! \left(
\frac{\kappa}{\bar{M}_5} \right)^{3/2} \!\! \bar{M}_{\mathrm{Pl}} \,
e^{-\pi \kappa r_c} \;,
\end{equation}
for all $m_n \ll \bar{M}_{\mathrm{Pl}} (\kappa/\bar{M}_5)^{3/2}$.
Here $x_n$ are zeros of the Bessel function $J_1(x)$.

The interactions of the gravitons with the SM fields on the TeV
brane are described by the effective Lagrangian
\begin{equation}\label{Lagrangian}
\mathcal{L}_{\mathrm{int}} = - \frac{1}{\bar{M}_{\mathrm{Pl}}} \,
h_{\mu\nu}^{(0)}(x) \, T_{\alpha\beta}(x) \, \eta^{\mu\alpha}
\eta^{\nu\beta} - \frac{1}{\Lambda_\pi} \sum_{n=1}^{\infty}
h_{\mu\nu}^{(n)}(x) \, T_{\alpha\beta}(x) \, \eta^{\mu\alpha}
\eta^{\nu\beta} \;,
\end{equation}
were $T^{\mu \nu}(x)$ is the energy-momentum tensor of the SM
fields, $h_{\mu\nu}^{(0)}$ is the field of the massless graviton.
The coupling constant of the KK gravitons is
\begin{equation}\label{Lambda_pi}
\Lambda_\pi \simeq \frac{\bar{M}_{\mathrm{Pl}}}{\sqrt{e^{2\Delta
\sigma} - 1}} \Big|_{\pi\!\kappa r_c \gg 1} \simeq
\bar{M}_{\mathrm{Pl}} \, e^{-\pi \kappa r_c} \;.
\end{equation}

The very expressions \eqref{m_n}, \eqref{Lambda_pi} do not depend on
$C$. Nevertheless, by taking different values of $C$ in
\eqref{sigma_solution}, we come to quite diverse \emph{physical
scenarios}. To demonstrate this, let us consider the following three
cases. From now on, it will be assumed that $\kappa > 0$ and $\pi
\kappa r_c \gg 1$.
\\

\noindent

1. $C = \kappa \pi r_c/2$. Then we get from \eqref{sigma_solution}
\begin{equation}\label{sigma_solution_1}
\sigma (y) = \frac{\kappa}{2} ( |y| - |y - \pi r_c | + \pi r_c) \;.
\end{equation}
It means that $\sigma_1 = 0$, $\sigma_2 = \kappa \pi r_c$. The
hierarchy relation looks like \cite{Randall:99}
\begin{equation}\label{hierarchy_relation_1}
\bar{M}_{\mathrm{Pl}}^2 = \frac{\bar{M}_5^3}{\kappa} \left(1 - e^{-2
\pi \kappa r_c} \right) \simeq \frac{\bar{M}_5^3}{\kappa}  \;.
\end{equation}
It requires $\bar{M}_5 \simeq \kappa \simeq \bar{M}_{\mathrm{Pl}}$.
Then the masses of the KK excitations are defined by eq.~\eqref{m_n}
to be
\begin{equation}\label{graviton_masses_1}
m_n  \simeq x_n \bar{M}_{\mathrm{Pl}} \, e^{-\kappa\pi r_c} \simeq
x_n \kappa \, e^{-\kappa\pi r_c} \;.
\end{equation}
Thus, the original RS1 model \cite{Randall:99} is realized in this
case. The KK spectrum of the model is a set of heavy resonances with
the lightest one around few TeV, if $\Lambda_\pi$ is chosen to be
about one TeV.
\\

\noindent

2. $C = 0$. This scheme was not yet considered by other authors. The
symmetry between the branes become very clear if a new variable $z =
y - \pi r_c/2$  is introduced, that results in%
\footnote{Under simultaneous replacements ($z \rightarrow - z$,
$\kappa \rightarrow - \kappa$), one gets $\sigma_1 \rightleftarrows
\sigma_2$ and $\sigma \rightarrow \sigma$.}
\begin{equation}\label{sigma_solution_2_z}
\sigma (z) = \frac{\kappa}{2} \left( \left|\frac{\pi r_c}{2} + z
\right| - \left|\frac{\pi r_c}{2} - z \right| \right) \;.
\end{equation}
Note, that $\sigma_1 = - \sigma_2 = -\kappa \pi r_c/2$. According to
eq.~\eqref{hierarchy_relation}, the hierarchy relation is given by
\begin{equation}\label{hierarchy_relation_2}
\bar{M}_{\mathrm{Pl}}^2 = \frac{2\bar{M}_5^3}{\kappa} \sinh (\pi
\kappa r_c)  \;,
\end{equation}
while the masses of the KK excitations are
\begin{equation}\label{graviton_masses_2}
m_n = x_n \kappa \, e^{-\kappa\pi r_c/2} \;.
\end{equation}
As for the coupling constant of the massive gravitons
\eqref{Lambda_pi}, it looks like
\begin{equation}\label{Lambda_pi_2}
\Lambda_\pi = \frac{\bar{M}_5^3}{\kappa \bar{M}_{\mathrm{Pl}}} \;.
\end{equation}
Then the hierarchy relation \eqref{hierarchy_relation_2} can be
rewritten as $\bar{M}_{\mathrm{Pl}} = 2 \Lambda_\pi \sinh (\pi
\kappa r_c)$.

Two different physical frameworks can be considered within scheme~2: \\
2a) $\bar{M}_5 = \kappa$. Then, according to \eqref{Lambda_pi_2},
the parameters of the model have the following large values
\begin{equation}\label{M5_2_a}
\bar{M}_5 = \kappa \simeq 4.9 \cdot 10^{10} \left(
\frac{\Lambda_\pi}{\mathrm{TeV}} \right)^{1/2} \mathrm{GeV} \;.
\end{equation}
The graviton masses are given by
\begin{equation}\label{m_n_a}
m_n = x_n \, \Lambda_\pi \;.
\end{equation}
For $\Lambda_\pi = 1$ TeV, one gets from the hierarchy relation
\eqref{hierarchy_relation_3} that $\kappa r_c \simeq 11.28$. Thus,
this scenario leads to a model with heavy graviton resonances, as in
the RS1 model.

2b) $\bar{M}_5 \gg \kappa$. This case leads to quite different
collider phenomenology. For instance, suppose that $\bar{M}_5 = 2
\cdot 10^{9}$ GeV and $\kappa = 10^4$ GeV. Then we get from
eq.~\eqref{Lambda_pi_2} that $\Lambda_\pi \simeq  3.3 \cdot 10^5$
GeV. The KK gravitons with the masses
\begin{equation}\label{m_n_2}
m_n = x_n \kappa \left( \frac{\Lambda_\pi}{\bar{M}_{\mathrm{Pl}}}
\right)^{1/2}
\end{equation}
form almost continuous spectrum,
\begin{equation}\label{m_n_2b}
m_n \simeq 3.7  \, x_n \, \mathrm{MeV} \;.
\end{equation}
Note that $\kappa r_c \simeq 9.43$ in this case.

The warp extra dimension can be searched for in the processes
mediated by $s$-channel exchanges of the KK gravitons.%
\footnote{For instance, $a\bar{a} \rightarrow h^{(n)} \rightarrow b
\bar{b}$, where $a/b = l, \gamma, q, g$, and $h^{(n)}$ is the KK
graviton.}
The \emph{universal} part of their matrix elements is defined by the
sum~\cite{Kisselev:06}
\begin{equation}\label{S_2}
\mathcal{S}(s) =  \frac{1}{\Lambda_{\pi}^2} \sum_{n=1}^{\infty}
\frac{1}{s - m_n^2 + i \, m_n \Gamma_n} \; ,
\end{equation}
where $\Gamma_n \simeq \eta \, m_n^3/\Lambda_{\pi}^2$ (with $\eta =
0.09$) denotes the total width of the graviton with the KK number
$n$ and mass $m_n$. By doing calculations analogous to those from
ref.~\cite{Kisselev:06}, with the use of relation \eqref{m_n}, we
obtain
\begin{equation}\label{S_2_solution}
\mathcal{S}(s) = - \frac{\bar{M}_5^3}{2\kappa^3 \Lambda_\pi^4}
\frac{1}{\sqrt{1 - \displaystyle 4 i \, \eta
\frac{s}{\Lambda_{\pi}^2}}} \left[ \frac{1}{z_1} \,
\frac{J_2(z_1)}{J_1(z_1)} - \frac{1}{z_2} \,
\frac{J_2(z_2)}{J_1(z_2)} \right],
\end{equation}
where
\begin{equation}\label{sigma_rho}
z_{1,2}^2 = \frac{1}{2i \eta} \left( \frac{\bar{M}_5}{\kappa}
\right)^3 \left[ 1 \mp \sqrt{1 - 4 i \eta \frac{s}{\Lambda_{\pi}^2}}
\right] \;.
\end{equation}

Taking into account that $|z_2| \gg |z_1| \gg 1$, we get
\begin{equation}\label{S_2}
\mathcal{S}(s) \simeq  - \frac{1}{2 \Lambda_\pi^3 \sqrt{s}} \left(
\frac{\bar{M}_5}{\kappa} \right)^{3/2} \frac{J_2(z_1)}{J_1(z_1)} \;,
\end{equation}
where
\begin{equation}\label{z_1}
z_1 \simeq \left( \frac{\bar{M}_5}{\kappa} \right)^{3/2}
\frac{\sqrt{s}}{\Lambda_\pi} \left[1  + \frac{i \eta}{2} \left(
\frac{\sqrt{s}}{\Lambda_\pi} \right)^2 \right] \;.
\end{equation}
Thus, in spite of the large value of the coupling constant
$\Lambda_\pi \simeq 330$ TeV, we come to a TeV physics, since
\begin{equation}\label{S_3_num}
|\mathcal{S}(s)| = \frac{F}{(1\mathrm{TeV})^3 \sqrt{s}} \;,
\end{equation}
where $F = \mathrm{O}(1)$ for our particular values of the
parameters $\bar{M}_5$, $\kappa$ and $\Lambda_\pi$.

By using asymptotic behavior of the Bessel functions, we obtain
\begin{equation}\label{KK_sum}
\mathcal{S}(s) = - \frac{1}{4 \Lambda_\pi^3 \sqrt{s}} \left(
\frac{\bar{M}_5}{\kappa} \right)^{3/2} \frac{\sin 2A + i \sinh (2 \,
\varepsilon)}{\cos^2 \! A + \sinh^2 \! \varepsilon} \;,
\end{equation}
where
\begin{equation}\label{parameters}
A = \frac{\sqrt{s}}{\Lambda_\pi} \left( \frac{\bar{M}_5}{\kappa}
\right)^{3/2}, \quad \varepsilon = \frac{\eta}{2} \Big(
\frac{\sqrt{s}}{\Lambda_\pi} \Big)^3 \!\! \left(
\frac{\bar{M}_5}{\kappa} \right)^{3/2} \;.
\end{equation}
For the chosen values of $\bar{M}_5$ and $\kappa$, we find that
\begin{equation}\label{parameters_numeric}
A \simeq 2.7 \cdot 10^5 \left( \frac{\sqrt{s}}{\mathrm{TeV}} \right)
\;, \quad \varepsilon \simeq 0.1 \Big( \frac{\sqrt{s}}{\mathrm{TeV}}
\Big)^3 \;.
\end{equation}

If an effective energy of colliding partons at the LHC $\hat{s}$ is
large enough, namely, $\sqrt{\hat{s}} \gtrsim 2.8$ TeV,
eq.~\eqref{KK_sum} can be significantly simplified,%
\footnote{At $\sqrt{s} = 2.8$ TeV, one gets $\sinh (2 \varepsilon)
\simeq 2 \sinh^2 \! \varepsilon \simeq 40$.}
\begin{equation}\label{zero_width_KK_sum}
\mathcal{S}_{\mathrm{asymp}}(\hat{s}) = - \frac{i}{2 \Lambda_\pi^3
\sqrt{\hat{s}}} \left( \frac{\bar{M}_5}{\kappa} \right)^{3/2} = -
\frac{i}{\sqrt{\hat{s}}} \left( \frac{\kappa \,
\bar{M}_{\mathrm{Pl}}^2}{\bar{M}_5^5} \right)^{3/2} \;.
\end{equation}
In contrast, at ILC energies ($\sqrt{s} \leqslant 500$ GeV) the
exact formula \eqref{KK_sum} should be used, since $\varepsilon \ll
1$  in this case, and a real part of $\mathcal{S}(s)$ is comparable
with (or larger than) an integer part.
\\

\noindent

3. $C = - \kappa \pi r_c/2$. In such a case,
eq.~\eqref{sigma_solution} means that
\begin{equation}\label{sigma_solution_3}
\sigma (y) = \frac{\kappa}{2} ( |y| - |y - \pi r_c | - \pi r_c) \;,
\end{equation}
and $\sigma_1 = -\kappa \pi r_c$, $\sigma_2 = 0$. Now the hierarchy
relation is of the form
\begin{equation}\label{hierarchy_relation_3}
\bar{M}_{\mathrm{Pl}}^2 = \frac{\bar{M}_5^3}{\kappa} \left( e^{2 \pi
\kappa r_c} - 1 \right) \simeq \frac{\bar{M}_5^3}{\kappa} \, e^{2
\pi \kappa r_c} \;.
\end{equation}
Correspondingly, the masses of the KK gravitons appear to be
proportional to the curvature $\kappa$
\begin{equation}\label{graviton_masses_3}
m_n = x_n \kappa \;.
\end{equation}

By using hierarchy relation \eqref{hierarchy_relation_3}, the
coupling constant of the massive gravitons on the TeV brane
\eqref{Lambda_pi} can be rewritten as
\begin{equation}\label{Lambda_pi_3}
\Lambda_\pi = \left( \frac{\bar{M}_5^3}{\kappa} \right)^{1/2} .
\end{equation}

Again, two diverse physical frameworks can be considered: \\
3a) To get $m_1 \sim 1$ TeV, we can put $\bar{M}_5 \sim \kappa \sim
1$ TeV. In such a case, one obtains a series of massive graviton
resonances in the TeV region which interact rather strongly with the
SM fields, since $\Lambda_\pi \sim 1$ TeV \cite{Kisselev:06}.

3b) More interesting scenario with the small curvature is realized,
if one takes $\kappa \ll \bar{M}_5 \sim 1$ TeV (RSSC model). For
instance, suppose that the fundamental gravity scale $\bar{M}_5$ is
of order few TeV, while the curvature $\kappa$ varies from hundreds
MeV to tens GeV. Then the graviton spectrum is almost continuous
\eqref{graviton_masses_3}, and it remains that in the model with one
flat extra dimensions \cite{Arkani-Hamed:98}-\cite{Arkani-Hamed:99}.
For the first time, this framework was considered in
refs.~\cite{Giudice:05}, \cite{Kisselev:05}. It was developed in our
forthcoming publications (see, for instance, \cite{Kisselev:06},
\cite{Kisselev:diphotons}, \cite{Kisselev:dimuons}).

In the limit of very small $\kappa$, when the inequality
\begin{equation}\label{small_cur_limit}
2\pi \kappa r_c \ll 1
\end{equation}
is satisfied, one get from \eqref{hierarchy_relation_3} the ADD-like
relation,
\begin{equation}\label{hierarchy_relation_small_cur}
\bar{M}_{\mathrm{Pl}}^2 = \bar{M}_5^3 (2\pi r_c) \;.
\end{equation}
At the same time, $\Lambda_\pi = \bar{M}_{\mathrm{Pl}}/\sqrt{2}$,
$m_n = n/r_c$\,. Nevertheless, it does not mean that RS-like scheme
with the small curvature $\kappa$ is equivalent to a 5-dimensional
space-time with one \emph{flat} ED, at least for not extremely small
values of $\kappa$. Indeed, as it follows from
\eqref{small_cur_limit}, \eqref{hierarchy_relation_small_cur},
\begin{equation}\label{cur_upper_limit}
\kappa \ll \frac{\bar{M}_5^3}{\bar{M}_{\mathrm{Pl}}^2} \;.
\end{equation}
Thus, the RSSC model can be though of as the theory with the flat ED
only if $\kappa < 10^{-22}$ eV, provided $\bar{M}_5 \sim 1$ TeV.
\\

It is often said that we have \emph{TeV physics} in the original RS1
model~\cite{Randall:99}, if the coupling constant $\Lambda_\pi$ is
about one or few TeV. Indeed, it is enough to put $\kappa r_c \simeq
10.54$ in \eqref{Lambda_pi} to obtain $\Lambda_\pi = 1$ TeV. But at
the same time, the hierarchy relation \eqref{hierarchy_relation_1}
requires $\bar{M}_5 \sim \kappa \sim
\bar{M}_{\mathrm{Pl}}$.%
\footnote{Correspondingly, the size of the extra dimensions is very
small, $r_c \simeq 50\, l_{\mathrm{Pl}}$.}

In the RSSC model a situation is completely different. The point is
that in the RSSC model a magnitude of all cross sections (after
summing up virtual or real gravitons) is defined by the fundamental
gravity scale $\bar{M}_5$, \emph{not by} $\Lambda_\pi$, provided
$\kappa \ll \bar{M}_5$~\cite{Kisselev:05}-\cite{Kisselev:06}. In
particular \cite{Kisselev:06},
\begin{equation}\label{S_3}
|\mathcal{S}(s)| \sim \frac{1}{\bar{M}_5^3 \sqrt{s}} \;.
\end{equation}
An analogous mechanism takes place in the ADD model
\cite{Arkani-Hamed:98}-\cite{Arkani-Hamed:99} in which the smallness
of the graviton coupling to the SM fields ($\sim
1/\bar{M}_{\mathrm{Pl}}$) is compensated by a huge number of
gravitons.

As a result, we come to the TeV physics, if $\bar{M}_5$ is about few TeV.%
\footnote{At the same time, $\Lambda_\pi \simeq 31.6 \,
(\bar{M}_5/\mathrm{TeV})^{3/2} (\mathrm{GeV}/\kappa)^{1/2}$ TeV can
be rather large.}
Thus, no new parameters of order $M_{\mathrm{Pl}}$ have to be
introduced in the RSSC scheme. At the same time, the hierarchy
relation \eqref{hierarchy_relation_3} is satisfied due to the large
warp factor $\exp(2\kappa\pi r_c)$. In particular, for $\bar{M}_5 =
1$ TeV and $\kappa = 100$ MeV, eq.~\eqref{hierarchy_relation_2} is
valid if $\kappa r_c \simeq 9.81$. For $\bar{M}_5 = 10$ TeV and
$\kappa = 1$ GeV, we get $\kappa r_c \simeq 9.08$.

For the time being, it is recognized that the hierarchy problem is
not solved in the ADD model
\cite{Arkani-Hamed:98}-\cite{Arkani-Hamed:99} in which the huge
value of the Planck scale is explained in terms of a new large
parameter which is a volume of extra dimensions. On the contrary, in
schemes 2 and 3 considered above the Planck mass is defined by the
large warp factors in the hierarchy relations
\eqref{hierarchy_relation_2}, \eqref{hierarchy_relation_3}.

Let us stress that the RS1 scenario (case 1) differs from other
scenarios, since it is the \emph{only} scheme with $\sigma_1=0$. As
a result, the coupling $\Lambda_\pi$ has no relation with the
parameters $\bar{M}_5$ and $\kappa$ in the limit $\pi\kappa r_c \gg
1$, but it is entirely defined by the warp factor \eqref{Lambda_pi}.
At the same time, $\bar{M}_{\mathrm{Pl}}$ depends very weakly on the
warp factor, as one can see from \eqref{hierarchy_relation_1}. On
the contrary, in the schemes with $\sigma_1 \neq 0$, the coupling
$\Lambda_\pi$ can be related to the model parameters $\bar{M}_5$ and
$\kappa$ via eqs.~\eqref{hierarchy_relation} and \eqref{Lambda_pi}
(see, for instance, eqs.~\eqref{Lambda_pi_2}, \eqref{Lambda_pi_3}).

Note that a shift $\sigma \rightarrow \sigma - B$, where $B$ is a
constant, is equivalent to a change of four-dimensional
coordinates~\cite{Rubakov:01},
\begin{equation}\label{x_transformation}
x^\mu \rightarrow x'^{\mu} = e^{-B} x^\mu \;.
\end{equation}
The effective gravity action on the TeV brane (with radion term
omitted) looks like (see, for instance, \cite{Boos:02})
\begin{equation}\label{gravity_TeV_action}
S_{\mathrm{eff}} = \frac{1}{4} \sum_{n=0}^{\infty} \int \! d^4x \!
\left[ \partial_\mu h^{(n)}_{\varrho\sigma}(x) \partial_\nu
h^{(n)}_{\delta \lambda} (x) \, \eta^{\mu\nu} - m_n^2
h^{(n)}_{\varrho\sigma} (x) h^{(n)}_{\delta \lambda}(x) \right] \!
\eta^{\varrho\delta} \eta^{\sigma\lambda} \;,
\end{equation}
The invariance of this action under transformation
\eqref{x_transformation} needs rescaling of the graviton fields and
their mass: $h^{(n)}_{\mu\nu} \rightarrow h'^{(n)}_{\mu\nu} = e^{B}
h^{(n)}_{\mu\nu}$, $m_n \rightarrow m'_n = e^{B} m_n$.

As an illustration, note that the transition from the RS1 model
(case 1) to the RSSC model (case 3) means that $\sigma \rightarrow
\sigma - \pi\kappa r_c$. Correspondingly,
eq.~\eqref{graviton_masses_1}
transforms into eq.~\eqref{graviton_masses_3}.%
\footnote{Of course, it does not mean that $m_n$ becomes larger in
the RSSC model, since very values of the parameter $\kappa$ are
quite different in these models.}



\section{Conclusions}

In the present paper the RS scenario with two branes is studied. We
used recently obtained general expression for the warp function
$\sigma(y)$ which has the explicit symmetry with respect to the
branes. The Einstein-Hilbert's equations define $\sigma(y)$ up to an
arbitrary constant $C$ \eqref{sigma_solution}. It was shown that,
depending on a value of $C$, one gets quite different physical
schemes. The well-known RS1 model can be realized as one particular
case. The other value of $C$ corresponds to the scheme with the
small curvature of the 5-dimensional space-time (RSSC model).
Contrary to the RS1 model, it has the almost continuous spectrum of
the KK gravitons. One more scheme is also suggested and studied
which can lead to an interesting collider phenomenology in the TeV
region.



\section*{Acknowledgements}

The author thanks profs. G.~Altarelli, I.~Antoniadis and
M.L.~Mangano for fruitful discussions. He is also indebted to the
Theory Division of CERN for the support and hospitality.



\end{document}